\pdfoutput=1

\documentclass[11pt]{article}

\usepackage{emnlp2021}

\usepackage{times}
\usepackage{latexsym}

\usepackage[T1]{fontenc}

\usepackage[utf8]{inputenc}

\usepackage{microtype}
\usepackage{pifont}
\usepackage{latexsym}

\usepackage{balance}
\usepackage{helvet}  
\usepackage{courier}  
\usepackage{url}  
\usepackage{graphicx}  
\usepackage{amssymb}
\usepackage{amsmath}
\usepackage{algorithm}
\usepackage{algorithmic}
\usepackage{amsthm}
\usepackage{multirow}
\usepackage{pgffor}
\usepackage{graphicx}
\usepackage{epstopdf}
\usepackage{tikz}
\usepackage{epstopdf}
\usepackage{mathtools}
\usepackage{enumitem}
\usepackage{subfigure}
\usepackage{tabulary}
\usepackage{color}
\usepackage{url}
\usepackage{flushend}
\usepackage{stfloats}
\usepackage{tcolorbox}
\usepackage[utf8]{inputenc}
\usepackage[english]{babel}
\usepackage{tabularx}
\usepackage{CJKutf8}
\usepackage{microtype}
\usepackage{textcomp}
\usepackage{booktabs}
\usepackage{colortbl}
\usepackage{soul}
\usepackage{arydshln}
\usepackage{natbib}
\usepackage{appendix}
\def\aclpaperid{226}
\usepackage{xspace}
\usepackage{microtype}
\usepackage{balance}
\usepackage{flushend}
\title{Dense Hierarchical Retrieval for Open-Domain Question Answering}
\author{
  \textbf{Ye Liu}$^1$\thanks{~~Work was done when the first author was a research intern at Salesforce Research}, \textbf{Kazuma Hashimoto}$^2$, \textbf{Yingbo Zhou}$^2$, \textbf{Semih Yavuz}$^2$, \textbf{Caiming Xiong}$^2$, \textbf{Philip S. Yu}$^1$ \\
$^1$ University of Illinois at Chicago, Chicago, IL, USA \\
$^2$ Salesforce Research, Palo Alto, CA, USA\\
  { \texttt{\{yliu279, psyu\}@uic.edu},} \\ {\texttt{\{yingbo.zhou, k.hashimoto, syavuz, cxiong\}@salesforce.com}}
}

\begin{document}
\maketitle
\begin{abstract}

Dense neural text retrieval has achieved promising results on open-domain Question Answering (QA), where latent representations of questions and passages are exploited for maximum inner product search in the retrieval process. However, current dense retrievers require splitting documents into short passages that usually contain local, partial and sometimes biased context, and highly depend on the splitting process. As a consequence, it may yield inaccurate and misleading hidden representations, thus deteriorating the final retrieval result. 
In this work, we propose Dense Hierarchical Retrieval (DHR), a hierarchical framework which can generate accurate dense representations of passages by utilizing both macroscopic semantics in the document and microscopic semantics specific to each passage. Specifically, a document-level retriever first identifies relevant documents, among which relevant passages are then retrieved by a passage-level retriever. The ranking of the retrieved passages will be further calibrated by examining the document-level relevance.
In addition, hierarchical title structure and two negative sampling strategies (i.e., \textit{In-Doc} and \textit{In-Sec} negatives) are investigated. We apply DHR to large-scale open-domain QA datasets. DHR significantly outperforms the original dense passage retriever, and helps an end-to-end QA system outperform the strong baselines on multiple open-domain QA benchmarks. 
\end{abstract}
\section{Introduction}
The goal of open-domain Question Answering (QA) is to answer a question without pre-specified source domain \citep{kwiatkowski2019natural}.
One of the most prevalent architectures in open-domain QA is the retriever-reader approach \citep{chen2017reading,lee2019latent}.
Given a question, the task of the retrieval stage is to identify a set of relevant contexts within a diversified large corpus (e.g., Wikipedia).
The reader component then consumes the retrieved evidence as input and predicts an answer.
In this paper, we focus on improving the efficiency and the effectiveness of the retrieval component, which in turn leads to improved overall answer generation for open-domain QA.

\begin{figure}[t]
\centering
\includegraphics[width=1\linewidth]{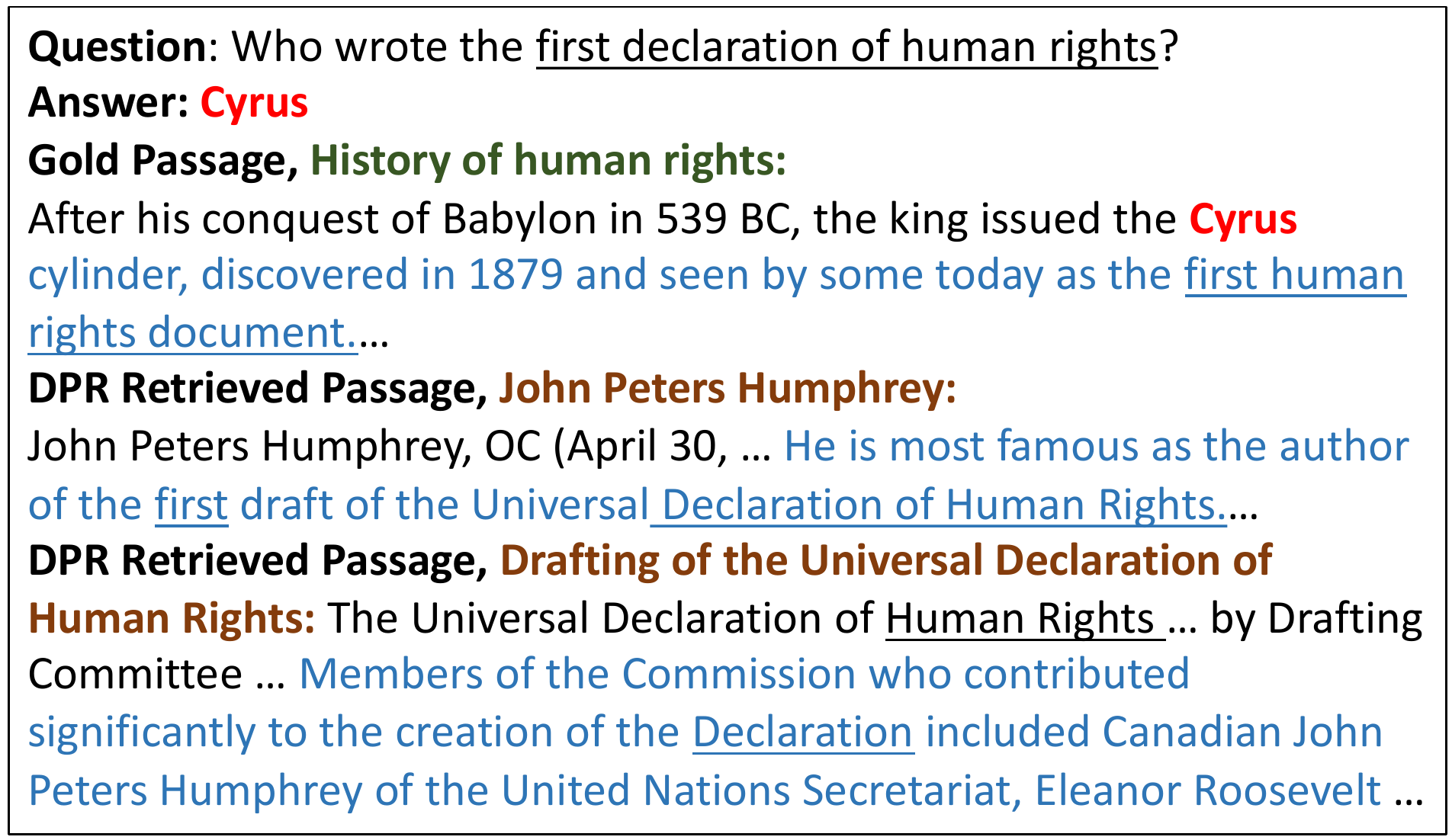}
\caption{An example of distracting passages in Natural Question \citep{kwiatkowski2019natural}. The first DPR retrieved passage shares similar semantics with the gold passage. The document title of the second DPR retrieved passage matches most question tokens. Both of the retrieved passages tend to result in a wrong answer.}
\label{figure:exp_intro}
\end{figure}

Pretrained transformer models, such as BERT~\citep{devlin2018bert}, are widely used in recent studies on the retriever-reader framework~\citep{asai2019learning,lewis2020retrieval,guu2020realm}.
To serve as input to the retriever, documents are split into short passages, and in the Dense Passage Retrieval, DPR~\citep{karpukhin2020dense}, a dual encoder framework is applied to encode questions and the split passages separately.
State-of-the-art dense retrievers outperform sparse term-based retrievers, like BM25 \citep{robertson2009probabilistic}, but they suffer from several weaknesses. 
First, due to the lack of effective pruning strategy, extracting relevant passages from a large corpus undergoes an efficiency issue especially in the inference time.
Second, given a question, many passages may comprehend similar topics with subtle semantic difference.
This fact requires the retriever and the reader to encode passages to their accurate semantic representations, which is an overwhelmed task.
Moreover, passages contain only local and specific information, thus easily leading to distracting representations.
As illustrated in Figure~\ref{figure:exp_intro}, distracting passages with similar semantics may lead to a wrong answer.

To alleviate these issues, we present a \textit{Dense Hierarchical Retriever} (DHR) framework, which consists of a dense document-level retriever and a dense passage-level retriever.
Document-level retriever aims at capturing coarse-grained semantics of documents in the sense that the embeddings of questions and their relevant documents are positively correlated.
The goal of document-level retriever is to prune answer-irrelevant documents and returns relevant ones, which will serve as a refined corpus to feed into passage-level retriever. 
Given relevant documents consisting of passages of similar topics, the passage-level retriever intends to identify the essential evidences, which may contribute to a correct answer.

In order to empower DHR, we formalize the hierarchical information of documents as a tree structure with two types of nodes, title nodes and content nodes. 
Then, a document is easily represented by its document summary, and a passage is represented by a hierarchical title list concatenated with its content. 
The benefit of using the hierarchical approach and exploiting the hierarchical information is three-fold: 
1) Coarse-grained information explicitly or implicitly covered in the document will guide the passage-level retriever to deviate from a fallacious embedding function; 
2) Passage-level retriever, a fine-grained component, will provide essential capability of identifying the necessary relevant evidences among similar passages; 
3) Document-level retriever prunes substantial amount of irrelevant and peripheral documents, and triggers a much faster inference. 
To further enhance the ability of the passage-level retriever in detecting gold passages among similar passages, we propose two negative sampling strategies (i.e., \textit{In-Doc} and \textit{In-Sec} negative sampling). 

Our main contributions are summarized as:
1) We propose a hierarchical dense retrieval on open-domain QA and achieve a fast inference speed with high retrieval precision;
2) The hierarchical information is used in a more structural way, which leads to a meaningful and global passage representation consistent with its document;
3) We conduct comprehensive experiments with state-of-the-art approaches on multiple open-domain QA datasets. 
Our empirical results demonstrate that we achieve comparable or better results in the open-retrieval setting.
Extensive ablation studies on various components and strategies are conducted.

\section{Notations and Preliminaries}
\label{sec:prelim}
\subsection{Text Retrieval for Open-Domain QA}
\label{sec:task_formu}
In open-domain QA, we are given a large corpus (e.g., Wikipedia) $\mathcal{C} = \left\{d_{1}, d_{2}, \ldots, d_{N}\right\}$, where each document $d_{i}$ is formed by a sequence of passages, $d_i=\{p^{(i)}_{1},p^{(i)}_{2},\ldots,p^{(i)}_{l}\}$.
The task of end-to-end open-domain QA can be formulated with a retriever-reader approach~\citep{chen2017reading}; we first find a passage (or a set of passages) relevant to a given question, and then use a reading comprehension model to actually derive its answer. 
It is common that we retrieve top-$k$ passages to be examined by the reading step.
The retrieval step is crucial, affecting the reading comprehension step.

\subsection{Dual Encoder Retrieval Model}
\label{sec:prelim_dual_encoder}
In the retrieval process, a commonly used approach referred as a dual encoder model \cite{bromley1993signature} consists of a question encoder $E_Q$ and a context encoder $E_P$, which encodes the question and the passage to $l$ dimensional vectors, respectively.
Unlike sparse term-based retrievers that rely on term frequency and inverse document frequency, dense neural retrievers formulate a scoring function between question $q$ and passage $p$ by the similarity of their embeddings, formalized as
\[
f_{\theta}(q,p)= \left\langle E_Q^{\theta}(q), E_P^{\theta}(p)\right\rangle,
\]
where $E_Q^{\theta}(q)\in \mathbb{R}^l$ and $E_P^{\theta}(p)\in \mathbb{R}^l$ are the embeddings, and $\langle \cdot,\cdot \rangle$ represents a similarity function such as doc product and cosine similarity.
Typically, $E_Q^{\theta}$ and $E_P^\theta$ are two large pre-trained models, e.g., BERT \cite{devlin2018bert}. 
We use different subscripts and same superscript $\theta$ to emphasize that these are two language models and fine tuned jointly.
DPR~\citep{karpukhin2020dense} is one of the representative models in this model family.

\noindent{\textbf{Contrastive Learning.}} Given a training set $\mathcal{S}=\{(q_1,y_1),\cdots,(q_m,y_m)\}$, we can create a training set $\mathcal{T}=\{(q_1,p_1^+,\mathcal{P}_1^-),\cdots,(q_m,p_m^+,\mathcal{P}_m^-)\}$ for the retrieval, where $q_i, y_i, p_i^+,\mathcal{P}_i^-$ are a question, its answer, its positive passage and a set of negative passages, respectively. 
All the selections of positive passages or documents in this paper are described in Appendix \ref{sec_training_example}.
For the set of negative passages $\mathcal{P}_i^-$, it is constructed in two ways: 1) BM25 negatives: top BM25-based passages not containing the answer; 2) In-batch negatives: passages paired with other questions appearing in the same mini batch. 

\noindent{\textbf{Training.}}
The objective of training is to learn an embedding function such that relevant pairs of questions and passages will have higher similarity than the irrelevant ones. For each training instance $(q_{i}, p_{i}^{+}, \mathcal{P}_{i}^{-})\in\mathcal{T}$, we contrastively optimize the negative log-likelihood of each positive passages against their negative passages, 
$$
\operatorname{loss}_\theta(q_{i}, p_{i}^{+}, \mathcal{P}_{i}^{-}) =-\log \frac{e^{f_{\theta}\left(q_{i}, p_{i}^{+}\right)}}{\sum_{p \in \{p_i^+\} \cup\mathcal{P}_{i}^{-}} e^{f_{\theta}\left(q_{i}, p\right)}}.
$$

\noindent{\textbf{Inference.}}
During inference, we encode the given question $q$ and conduct the maximum inner product search between $E_Q^\theta(q)$ and $E_P^\theta(p)$ for every passage $p$. Then a ranked list of k most relevant passages are served as input of the reader.

\section{Dense Hierarchical Retrieval (DHR)}

\begin{figure}[t]
\centering
\includegraphics[width=0.98\linewidth]{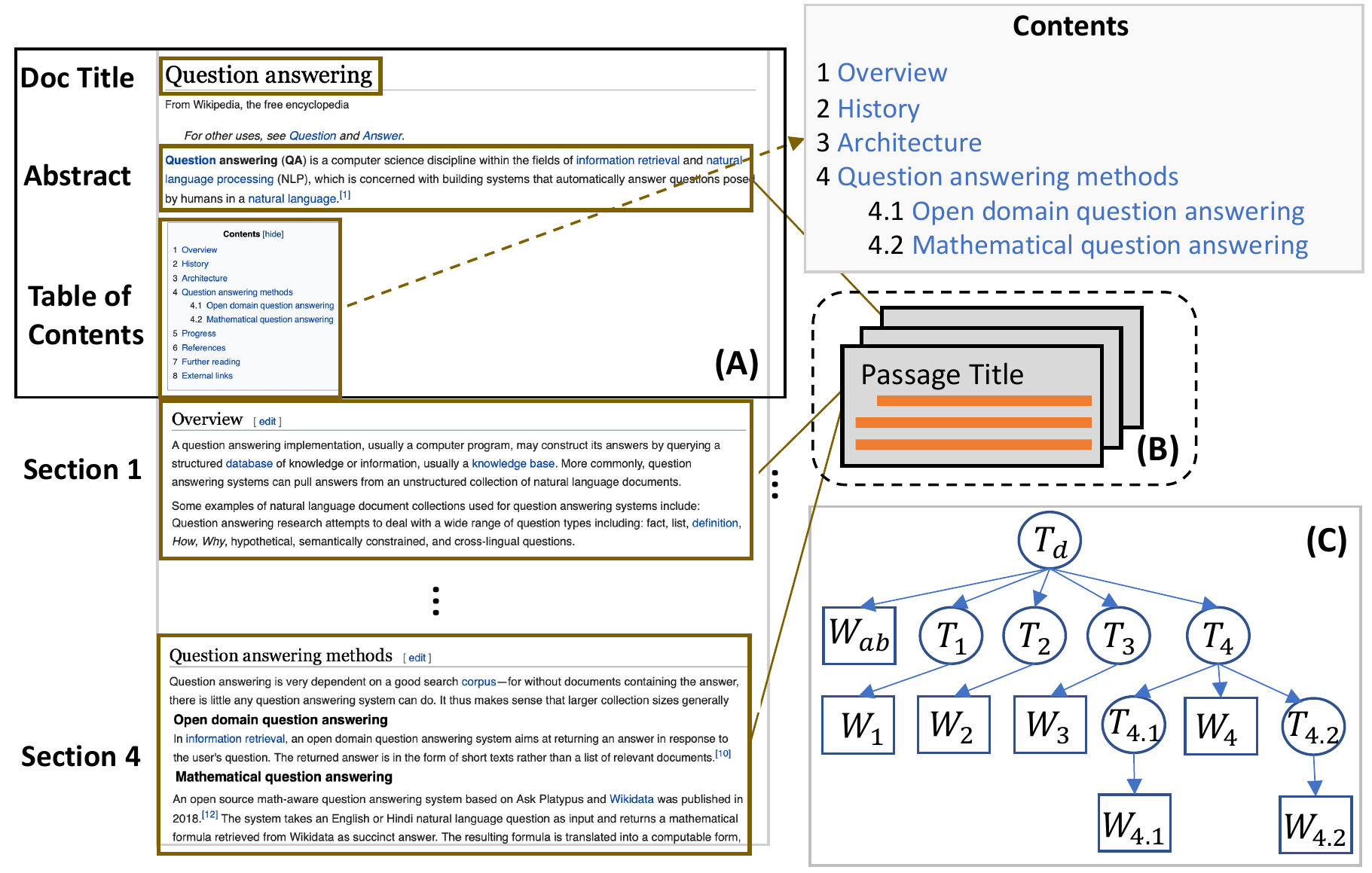}
\caption{An illustration of a typical Wikipedia page. (A) Document representation. (B) Passage representations. (C) Hierarchical title structure, i.e., title tree.}
\label{fig_wiki_intro}
\end{figure}

\begin{figure*}[t]
\centering
\includegraphics[width=0.98\linewidth]{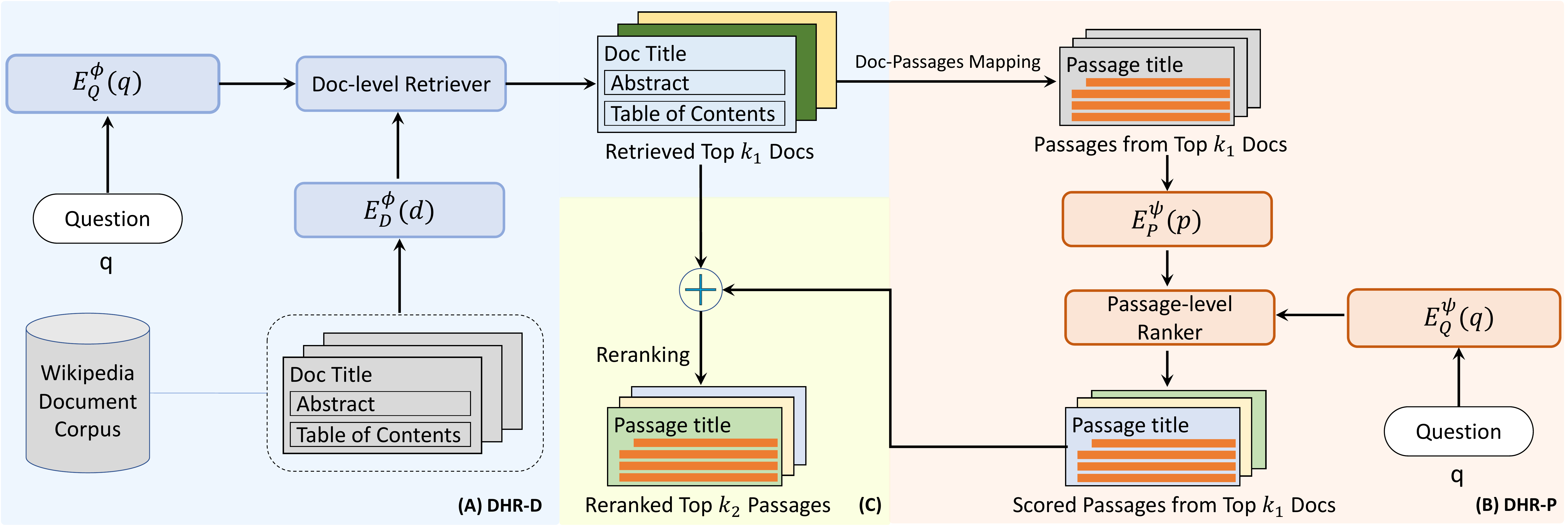}
\caption{An overview of DHR. During inference, document-level retriever first retrieves top-$k_{1}$ documents (as shown in (A)). Then, passage-level retriever scores the passages in top-$k_{1}$ documents (as shown in (B)). At last, DHR reranks passages based on two levels of relevance scores and return top-$k_{2}$ passages (as shown in (C)).}
\label{figure:method}
\vspace{-2mm}
\end{figure*}

This section presents our Dense Hierarchical Retrieval (DHR) model, which consists of a Dense Document-level Retrieval (DHR-D) and a Dense Passage-level Retrieval (DHR-P).
Figure~\ref{figure:method} shows an overview of our proposed method.

\subsection{Structural Document}
\label{sec_struc_doc}
A structured web article like a Wikipedia page in Figure~\ref{fig_wiki_intro} contains a document title, abstract, table of contents and different levels of sections consisting of titles and paragraphs.
To better leverage the hierarchical information of the document, we formalize the structural document as a tree structure called title tree with the hierarchical title structure being the backbone.
The title tree uses the document title $T_d$ as the root, the section titles of different levels as intermediate nodes, and the textual content under the same title as a leaf.
Note that there are two types of nodes namely title node and content node.
Each title or content will appear in the tree exactly once.

\subsection{Dense Document-level Retrieval (DHR-D)}
\label{sec_dhrd}
Dense Document-level Retrieval (DHR-D) aims at capturing the semantics of the documents in the sense that the embeddings of the questions and their relevant documents are positively correlated. DHR-D employs a BERT-based dual encoder model consisting of a question encoder $E_Q^\phi$ and a document encoder $E_D^\phi$, where $\phi$ emphasizes that two encoders are trained jointly. The relevance score of a document to a question is computed by dot product of their dense representation vectors:
\begin{equation}
f_\phi\left(q, d\right)=\langle \operatorname{E}_{Q}^\phi(q),~ \operatorname{E}_{D}^\phi\left(d\right)\rangle,
\end{equation}
where $\operatorname{E}_{Q}^\phi(q)\in \mathbb{R}^l$, $\operatorname{E}_{D}^\phi(d)\in \mathbb{R}^l$ and $\langle\cdot\rangle$ represents the dot product. 

\noindent{\textbf{Document Representation.}}
In order to enable the document encoder to capture holistic view of the documents covering their essential topics \cite{chang2020pre}, we use their summary as input to $E_D^\phi$.
We define \textit{document summary} as a concatenation of title $T_d$, abstract $W_{ab}$, and the linearized table of contents $T_{table}$.
We linearize the table of contents by following a pre-order traversal on only title nodes of the title tree excluding the root node $T_d$.
Separating each title by the special token $\texttt{[SEP]}$ (or comma), we finalize the representation of table of contents as
$T_{table} = T_{1}$ \texttt{[SEP]} $T_{1.1}$ \texttt{[SEP]} $\cdots$ \texttt{[SEP]} $T_{i.\cdots.l}$. Then the final representation of the document summary to be consumed by $E_D^\phi$ is defined as $d=$\texttt{[CLS]} $T_{d}$ \texttt{[SEP]} $W_{ab}$ \texttt{[SEP]} $T_{table}$ \texttt{[SEP]}. 

\noindent{\textbf{Negative Sampling.}}
Recall that given a training sample $(q_i,y_i)\in\mathcal{S}$ for open-domain QA, we can create a contrastive training instance with $(q_i,d_i^+,\mathcal{D}_i^-)$ for the retrieval, where $q_i, y_i, d_i^+,\mathcal{D}_i^-$ correspond to question, answer, positive document and a set of negative documents, respectively. With respect to the negative documents, besides leveraging in-batch negatives similar to DPR \cite{karpukhin2020dense}, we introduce two negative document sampling strategies: 
1) \textit{Abstract negatives}: we select top-ranked documents by BM25 whose abstract contains most question tokens but the whole document content doesn't contain the gold answer;
2) \textit{All-text negatives}: we select top-ranked documents by BM25 whose whole document content contains most question tokens but doesn't contain the gold answer. 

We train DHR-D following the strategy in Section \ref{sec:prelim_dual_encoder}.
We also optimize the negative log-likelihood. The only difference here is that the examples are documents instead of passages. During inference, DHR-D extracts a set of relevant documents for each question which then serve as a more focused evidence pool to be processed by Dense Passage-level Retrieval (DHR-P). The overall inference process will be elaborated in Section \ref{sec_inference}.

\subsection{Dense Passage-level Retrieval (DHR-P)}
\label{sec_dhrp}
Given relevant documents from DHR-D, the goal of our Dense Passage-level Retrieval (DHR-P) is to detect the most crucial evidence that may contribute to answer the question. Another dual encoder model is used in DHR-P with one BERT representing the question encoder $E_Q^\psi$ and the other BERT representing the passage encoder $E_P^\psi$. Notice that we use $\psi$ here to distinguish between the pair of encoders used for DHR-P and DHR-D. 
Similarly, the relevance score between passage $p$ and question $q$ is calculated by the dot product of their semantic embeddings:
\begin{equation}
f_\psi\left(q, p\right)=\langle \operatorname{E}_{Q}^\psi(q),~ \operatorname{E}_{P}^\psi\left(p\right)\rangle.
\end{equation}

\noindent{\textbf{Passage Representation.}} 
Here we describe two major differences of our passage representation from the previous work. 
First, instead of naively splitting document into passages, we only allow the passages within the same section to split. 
In other words, splitting can only happen within each leaf in the title tree. 
In this way, each passage will be semantically more consistent. 
Second, to magnify the differences between passages of similar topics, we augment each passage with a passage title. 
We define a \textit{passage title} as the concatenation of titles on the path from the root node to its content leaf using special token \texttt{[SEP]} (or comma) as the separator inserted in between.
Then the passage $p$ can be represented as \texttt{[CLS]} $T_{d_p}$ \texttt{[SEP]} $T_{i_1}$ \texttt{[SEP]} $T_{i_2}$ \texttt{[SEP]} $\cdots$ \texttt{[SEP]} $T_{i_n}$ \texttt{[SEP]} $W_{p}$ \texttt{[SEP]}, where $d_p$ represents the document it belongs to and $W_{p}$ indicates the passage content.

\noindent{\textbf{Negative Sampling.}} In order to improve the capability of detecting the essential evidence among similar passages, we propose two hard negative sampling strategies for DHR-P. 
Besides using BM25 negatives and in-batch negatives,  we propose \textit{In-Doc negative} and \textit{In-Sec negative}. While \textit{In-Sec} negatives are the passages which are in the same section with the positive passage but don't contain the answer. \textit{In-Doc} negatives are passages which are in the same document with the positive passage but don't contain the answer. 

\subsection{Iterative Training}
\label{sec_iter_train}
Inspired by improvement of using semantically related negative examples generated by previous checkpoint in ANCE \cite{xiong2020approximate}, we adopt an iterative training scheme for both DHR-D and DHR-P. 
More precisely, we use the retriever resulting from the initial phase of training to generate hard negative examples, which may be semantically quite related to the question but don't contain the answer.
From the perspective of adversarial training \cite{madry2017towards}, these negative examples can also be regarded as adversarial examples. 
Hence, training on these examples will increase the robustness of the models. 
Therefore, in the second iteration we further train DHR-D and DHR-P with generated negative examples together with the negative examples used in the first iteration. 

\subsection{Inference}
\label{sec_inference}
Before inference, we apply the document encoder $\operatorname{E}_{D}^\phi$ to all the documents and index them using efficient FAISS \cite{johnson2019billion} offline. Given a question $q$ at inference time, we compute its embeddings $\operatorname{E}_{Q}^\phi(q)$ and $\operatorname{E}_{Q}^\psi(q)$ for DHR-D and DHR-P respectively. Then, we first retrieve top-$k_1$ relevant documents using the index built by DHR-D. 
The passages of the top-$k_1$ retrieved documents then serve as a refined corpus, upon which $\operatorname{E}_{P}^\psi$ is applied to select top-$k_2$ passages.

Although DHR-D has already helped pruning of irrelevant documents before passage retrieval, we still leverage document-level similarity score it offering in addition to passage similarity score for the final ranking.
Thus, we define the final passage score function as a combination of relevance scores provided by DHR-D and DHR-P:
\begin{equation}
    f(q,p_i)= f_\psi\left(q, p_{i}\right) + \lambda \cdot f_\phi\left(q, d_{p_i}\right)
\label{eq_coef}
\end{equation}
where $d_{p_i}$ is the document $p_i$ belongs to and $\lambda$ is the coefficient controlling the balance. 
We noticed that the relevance scores of DHR-D and DHR-P are in the close scale for all our experiments. $\lambda\in[0.5,1]$ is a quite robust choice for desired performance.

\section{Experimental Setup}
In this section, we describe the dataset and basic setup for experiments.

\subsection{Wikipedia Data Pre-processing} \label{subsection:wiki_processing}
Following \citet{karpukhin2020dense}, we use the English Wikipedia dump from Dec. 20, 2018 as the source documents for answering questions. We first apply the WikiExtractor to extract the clean, textual documents with hierarchical title list from the Wikipedia dump, which removes semi-structured data, such as tables, infoboxes, lists, as well as disambiguation pages. 
In DPR \cite{karpukhin2020dense}, all the texts under the same document are first concatenated as a single block, which is then split into multiple blocks of fixed-length passages of 100 words, discarding the blocks of shorter than 100 words. 
We follow a different, more principled strategy to avoid ending up with abruptly broken and unnatural passages.
To this end, we concatenate the text under the same section and split each section into multiple, disjoint text blocks, whose maximum length is not over 100 words. 
Following this strategy, we obtain 25,992,490 passages from 5,380,681 documents.

\begin{table}[htp]
\centering
\resizebox{0.49\textwidth}{!}{
\begin{tabular}{lccccc}
\toprule
\textbf{Dataset}           & \multicolumn{2}{c}{\textbf{Train}} & \multicolumn{2}{c}{\textbf{Dev}} & \textbf{Test} \\ \hline
NQ &   79,168  & 59,906            &  8,757 & 6,610 &  3610    \\
TriviaQA &   78,785 &   60,314          &   8,837 & 6,753 &   11,313   \\
WebQuestions   &    34,17 &  2,432           &  361 & 275  &    2,032  \\
CuratedTrec   &  1,353   &  1,114     &  133 & 114  &    694  \\
\bottomrule
\end{tabular}
}
\caption{Number of questions in each QA dataset. The two columns of \textbf{Train} and \textbf{Dev} denote the original examples in the dataset and the actual questions used. See Section \ref{sec_training_example} for more details.}
\label{Table_QA_dataset}
\end{table}

\subsection{Question Answering Datasets}
We use four QA datasets that have been most commonly used benchmarks for open-domain QA evaluation \cite{lee2019latent,karpukhin2020dense}: \\
\textbf{Natural Questions (NQ)}~\citep{kwiatkowski2019natural}
consists of questions mined from real Google search queries, for which the answers are spans in Wikipedia documents identified by annotators. \\ 
\textbf{TriviaQA}~\citep{joshi2017triviaqa} contains a set of trivia questions with answers that were originally scraped from the Web. \\
\textbf{WebQuestions}~\citep{berant2013semantic} consists of questions selected using Google Suggest API, where the answers are entities in Freebase.\\
\textbf{CuratedTREC (TREC)}~\citep{baudivs2015modeling} is a collection of questions from TREC QA tracks as well as various Web sources, intended for open-domain QA from unstructured text.

\subsection{Selection of positive documents and passages.} \label{sec_training_example}
To determine the positive passage (or document), we assign it as the passage (or document) containing the gold context of the answer when it is given by human annotation; otherwise we feed the question to a BM25 system to retrieve the top-1 passage (or document) containing the answer as the positive passage (or document).

For Natural Questions, since the relevant context and document title are provided, we directly use the provided document title to find our processed corresponding document and use the relevant context map to our processed passage in the candidate pool. The questions are discarded when the matching is failed due to different Wikipedia versions or pre-processing. Because only pairs of questions and answers are provided in TREC and TriviaQA, we use the highest-ranked passage from BM25 that contains the answer as the positive passage and its belonging document as the gold document. If none of the top 100 retrieved passages has the answer, the question will be discarded. 
For WebQ, since it contains the gold title in the Freebase, we try both ways (matching and BM25 ranking) and find that using the highest-ranked passage from BM25 as the positive passage and its belonging document as the positive document can produce better performance than using the gold title from Freebase. We think it is due to the discrepancy between Freebase and Wikipedia corpus. 
Table \ref{Table_QA_dataset} shows the number of questions in training/dev/test sets for all the datasets and the actual questions used in training and dev sets.

\section{Experiments}
In this section, we evaluate the performance of our Dense Hierarchical Retriever (DHR\footnote{\url{https://github.com/yeliu918/DHR}}), along with analysis on how each component affects the results and the retrieval time efficiency. For the retrieval implementation detail, please refer to Appendix \ref{sec_retrieval_implementation}.

\begin{table*}[tp]
\centering
\resizebox{1\textwidth}{!}{
\begin{tabular}{l|ccc|ccc|ccc|ccc}
\toprule
\multirow{2}{*}{\textbf{Retriever}}   & \multicolumn{3}{c|}{\textbf{NQ}} & \multicolumn{3}{c|}{\textbf{TriviaQA}}  & \multicolumn{3}{c|}{\textbf{WebQ}} & \multicolumn{3}{c}{\textbf{TREC}}\\
 & \textbf{Top-1} & \textbf{Top-20} & \textbf{Top-100} & \textbf{Top-1} & \textbf{Top-20} & \textbf{Top-100} & \textbf{Top-1} & \textbf{Top-20} & \textbf{Top-100} & \multicolumn{1}{c}{\textbf{Top-1}} & \multicolumn{1}{c}{\textbf{Top-20}} & \multicolumn{1}{c}{\textbf{Top-100}} \\ \hline
BM25  & -  & 59.1& 73.7 & -  & 66.9& 76.7 & -  & 55.0& 71.1 &  & 70.9  & 84.1   \\
BM25*  & 18.48     & 60.19      & 75.98& 40.02     & 72.78      & 81.03& 17.03     & 56.45      & 73.77& 27.89& 74.97 & 87.92  \\ \hline
\textbf{\textit{1-iter}} & & & & & & & & & & \\ 
DPR    & 45.87     & 79.97      & 85.87& -  & 79.4& 85.0 & -  & 73.2& 81.4 & - & 79.8  & 89.1   \\
DPR*   & 40.08     & 81.05      & 88.31& \underline{52.94}     & \underline{80.43}      & \underline{85.40} & \underline{35.78}     &\underline{73.87}      & \underline{81.64}& \underline{36.03}& \underline{80.70} & \underline{90.49}  \\ 
\cite{lu2020neural} &  \underline{52.0}   & \underline{82.8} & \underline{88.4} & - & - & - & - & - &- &- &- & -\\ \hdashline
DHR & \textbf{55.37} & \textbf{85.07}  & \textbf{89.92}   & \textbf{54.40}   & \textbf{80.81}    & \textbf{85.69}   &  \textbf{36.86}  &  \textbf{73.98}   &  \textbf{82.69}    & \textbf{48.27} & \textbf{84.01}  &  \textbf{91.26}  \\\hline
\textbf{\textit{2-iter}} & & & & & & & & & &\\ 
DPR    & 52.47     & 81.33      & 87.29& -  & -   & -    & -  & -   & -    & -  & -   & -    \\ 
DPR*   & \underline{52.67}     & \underline{84.67}      & \underline{89.95}& \underline{53.89}   & \underline{79.68}   &  \underline{85.63}    & \underline{38.44}     & \underline{75.19}      & \underline{82.87}& \underline{41.35}& \underline{79.68} & \underline{91.21}  \\  \hdashline
DHR    & \textbf{57.04} & \textbf{85.60}  & \textbf{90.64}   & \textbf{55.08}   & \textbf{80.76}    &  \textbf{85.97}   &  \textbf{41.73} &  \textbf{75.29}   &   \textbf{83.05}   &  \textbf{48.42}  & \textbf{84.17}  &  \textbf{91.34}  \\
\bottomrule
\end{tabular}
}
\caption{Top-1, Top-20 and Top-100 passage-level retrieval accuracy on test sets, measured as the percentage of top 1/20/100 retrieved passages that contain the answer. * represents the reproduction on our processed Wikipedia data. We bold the best performance and underline the second best performance.
}
\label{Table_DHR_result}
\end{table*}

\subsection{Main Results}
In Table \ref{Table_DHR_result}, we report the retrieval performance of different systems on four QA datasets in terms of Top-1, 20 and 100 passage-level retrieval accuracy.  
For a fair comparison, we first re-implemented the DPR method on the Wikipedia data processed with our passage construction strategy defined in Section \ref{subsection:wiki_processing}, which is denoted as DPR* in Table \ref{Table_DHR_result}.
The retrieval performance of DPR* outperforms the original DPR on all datasets, except the Top-1 retrieval performance on NQ, showing the clear advantage of using our more principled in-section splitting strategy, which can better preserve the contextual consistency in each passage. 
Secondly and perhaps most importantly, we would like to highlight the benefit of our proposed hierarchical dense retrieval method (DHR) over the baseline DPR*.
As shown in Table \ref{Table_DHR_result}, DHR consistently and significantly outperforms DPR* across the board over four datasets we conduct experiments on. 
Most notably, it can provide up to 12\% and 4\% absolute improvement in top-1 and top-20 retrieval accuracy over DPR* on NQ and TREC benchmarks.
Also, we observe that DHR's improved retrieval performance translates well on to the iterative training setting. 
More precisely, using the wrong passages that are found semantically relevant by the first iteration model as negatives for training the second iteration greatly helps further improve the performance of DHR. 
Although the iterative training significantly boosts the performance of DPR*, our proposed DHR model still significantly outperforms DPR* across the four datasets, which is consistent with the conclusion from the first iteration setting.
Finally, we note that DHR also improves upon a recent work \cite{lu2020neural}, which achieves significant improvement over DPR using better negative samples.





\subsection{Ablation Study}
To further understand how each component of DHR works, we conduct several additional experiments on both Document-level retrieval and Passage-level retrieval on the NQ dataset.

\noindent{\textbf{Ablation Study on DHR-D.}}
From Table \ref{Table_DHR-D}, our Doc-level retrieval accuracy greatly outperforms the result of BM25, which shows the efficiency of our dense document-level retriever. Comparing the retriever results of using \textit{Abstract negative} with \textit{All-text negative} in lines 2 and 3, using \textit{Abstract negative} outperforms the performance of using \textit{All-text negative}, which may due to the noisy context bringing from the whole document context harming the performance.  

We test the influence of whether uses the table of contents as the context in the document-level retriever. As shown in the bottom block of Table \ref{Table_DHR-D}, the table of contents can improve the performance of document-level retrieval considerably, which demonstrates our assumption that the table of contents can be viewed as the highlight or summarization of the document contents.

\begin{table}[tp]
\centering
\resizebox{0.49\textwidth}{!}{
\begin{tabular}{lcccc}
\toprule
\textbf{Retriever}  & \textbf{Top-1} & \textbf{Top-5} & \textbf{Top-20} & \textbf{Top-100} \\ \hline
BM25  & 28.95 & 54.21 & 71.97  & 83.88   \\ \hdashline
DHR-D(Abs)                 & \underline{65.32} & \underline{82.85} & \underline{88.75}  & \underline{92.35}   \\
DHR-D(All)                & 64.04 & 81.81 & 87.71  & 92.11   \\ \hdashline
DHR-D(Abs)+T        & \textbf{68.28} & \textbf{83.80} & \textbf{89.28}  & \textbf{92.83}   \\ \hline
\textit{2-iter}  & & & &\\
DHR-D(Abs)+T  & \textbf{71.86} & \textbf{85.35} & \textbf{90.30}  & \textbf{93.16}   \\
\bottomrule
\end{tabular}
}
\caption{Top-1, Top-5, Top-20 and Top-100 document-level retrieval accuracy on NQ test sets. Abs denotes \textit{Abstract Negative}. All means \textit{All-text Negative}. T means that we add the table of contents into the document context.}
\label{Table_DHR-D}
\end{table}

\noindent{\textbf{Ablation Study on DHR-P.}}
To fairly compare with DPR, all the results of DHR-P in this section are from retrieving the whole passage corpus without the help of DHR-D to retrieve the relevant documents. 
We introduced two different ways to linearize the passage title tree in Table \ref{Table_DHR-P}.
The comparison results of Tc and Tt show that using a comma as a separator is better than using a special token \texttt{[SEP]}, and containing the passage title with the passage context is better than without it.
We think it shows that the hierarchical passage title can help the passage context capturing more global information from the document and help the retriever achieve better performance.     

As shown in the bottom block of Table \ref{Table_DHR-P}, the \textit{In-Doc negative} and \textit{In-Sec negative} improve the passage-level retrieval accuracy, which verifies the idea that improving the passage-level retrieval to distinguish the positive passage from the other passages in the same document is a simple and effective way. The reason why \textit{In-Doc negative} outperforms \textit{In-Sec negative} is that the number of the passage in the same section is less than the number of the same document passage and the passages in the same document also share the close semantic similarity.  


\begin{table}[tp]
\centering
\resizebox{0.49\textwidth}{!}{
\begin{tabular}{lcccc}
\toprule
\textbf{Retriever}  & \textbf{Top-1} & \textbf{Top-5} & \textbf{Top-20} & \textbf{Top-100} \\ \hline
DPR*    & 40.08 & 66.79 & 81.05  & 88.31   \\ \hdashline
DHR-P+Tc& 43.74 & 68.67 & 81.42  & 88.75   \\
DHR-P+Tt& 43.67 & 68.39 & 81.05  & 88.81   \\ \hdashline
DHR-P(Sec)+Tc & 50.17 & 71.80 & 82.16  & 88.12   \\
DHR-P(Doc)+Tc & \textbf{51.61} & \textbf{73.16} & \textbf{82.87}  & \textbf{89.16}   \\ \hline
\textit{2-iter} & & & & \\
DHR-P(Sec)+Tc & 54.46 & 75.54 & 84.99  & 90.19   \\ 
DHR-P(Doc)+Tc  & \textbf{55.12} & \textbf{76.06} & \textbf{85.01}  & \textbf{90.19}   \\
\bottomrule
\end{tabular}
}
\caption{Top-1, Top-5, Top-20 and Top-100 passage-level retrieval accuracy on NQ test sets. Tc and Tt denote using comma and \texttt{[SEP]} to separate the passage title, respectively. Sec denotes using \textit{In-Sec negative}. Doc means using \textit{In-Doc negative}.}
\label{Table_DHR-P}
\end{table}

\noindent{\textbf{Ablation Study on Reranking.}}
In Table \ref{Table_DHR}, we use the DHR-P model with In-Doc negative and title, which achieves the best performance in Table \ref{Table_DHR-P} on the whole passage corpus to compare with the two-step hierarchical retrieval models. DHR w/o rerank denotes the passage-level retrieval result from the passage corpus of Top-$k_{1}$ relevant documents without reranking. DHR w/o rerank outperforms DHR-P, demonstrating that the Doc-level retrieval can eliminate the distracting documents which could harm the Passage-level retrieval. 

We propose different ways to combine the Doc-level and Passage-level retriever scores to rerank the passages. DHR w rerank is the serial strategy proposed in the paper that we first use the Doc-level retriever to get the Top-$k_{1}$ relevant documents and use the Passage-level ranker to score the passage from the retrieved documents and rerank them based on the combination of Doc-level and Passage-level similarity scores. DHR para rerank is a parallel way to rank the passages. Firstly, Doc-level retriever scores all documents and Passage-level retriever scores all passages in the corpus. Then the model aggregates those two scores together for each question. 
The result in Table \ref{Table_DHR} shows the effectiveness of our approach and demonstrates that using the Doc-level retriever first to limit the documents to a small relevant set will not harm the overall retrieval performance but help filter out some answer-irrelevant documents.
Moreover, both the rerank methods outperform DHR w/o rerank, which shows the necessity of the reranking. 


\begin{table}[tp]
\centering
\resizebox{0.47\textwidth}{!}{
\begin{tabular}{lcccc}
\toprule
& \textbf{Top-1} & \textbf{Top-5} & \textbf{Top-20} & \textbf{Top-100} \\ \hline
DHR-P(Doc)+Tc     & 51.61 & 73.16 & 82.87  & 89.16   \\
DHR w/o rerank  & 52.80 & 73.82 & 83.80  & 89.81   \\
DHR w rerank & \textbf{55.68} & \textbf{75.51} & \textbf{84.96}  & \textbf{89.85}   \\
DHR para rerank & \underline{55.29} & \underline{75.10} & \underline{84.24} & \underline{89.15}   \\
\midrule
\textit{2-iter} & & & & \\
DHR-P(Doc)+Tc & 55.12 & 76.06 & 84.99  & 90.19   \\
DHR w/o rerank  & \underline{55.90} & \underline{76.32} & \underline{85.18}  & \underline{90.42}   \\
DHR w rerank  & \textbf{56.62} & \textbf{76.54} & \textbf{85.35}  & \textbf{90.53}   \\
\bottomrule
\end{tabular}
}
\caption{Top-1, Top-5, Top-20 and Top-100 passage-level retrieval accuracy on NQ test sets. DHR para rerank represents the parallel generating the document-level and passage-level similarity scores and add them based on Eq. \ref{eq_coef}.}
\label{Table_DHR}
\end{table}

\subsection{Hyperparameter Sensitivity Analysis} \label{sen_anly}
\begin{table}[tp]
\centering
\resizebox{0.47\textwidth}{!}{
\begin{tabular}{lcccc}
\toprule
& \textbf{Top-1} & \textbf{Top-5} & \textbf{Top-20} & \textbf{Top-100} \\ \hline
\textit{1 iter}  & & & & \\
DHR($\lambda$=1) & 55.68 & 75.51 & 84.96  & 89.85 \\
DHR($\lambda$=0.57) & \textbf{55.37} & \textbf{75.43} & \textbf{85.07}  & \textbf{89.92}  \\
\midrule
\textit{2-iter} & & & & \\
DHR($\lambda$=1)  & 56.62 & 76.54 & 85.35  & 90.53   \\
DHR($\lambda$=0.50) & \textbf{57.04} & \textbf{77.06} & \textbf{85.60}  & \textbf{90.64}  \\
\bottomrule
\end{tabular}
}
\caption{Top-1, Top-5, Top-20 and Top-100 passage-level retrieval accuracy on NQ test sets.}
\label{Table_lambda}
\end{table}


We analyze the parameter $\lambda$, which is used in Eq. \ref{eq_coef} as the coefficient of combining doc-level score and passage-level score. We tuned the $\lambda$ values on different datasets by optimizing Top-20 retrieval accuracy on the development set. We obtained the optimal weight by performing a grid search in the range [0, 2]. We started with step size 0.1 and found the optimal $\lambda_{1}$. Then, we used step size 0.01 in the range $[\lambda_{1}-0.05, \lambda_{1}+0.05]$ to find the optimal $\lambda$.
From the results in Table \ref{Table_lambda}, we can see that directly adding two scores together ($\lambda$=1) can lead to the good performance compared with the best performance model ($\lambda$=0.57 first iter, $\lambda$=0.5 second iter), which shows the robustness of the model without too many parameters tuning. 

For the top-$k_{1}$ retrieved documents that are given to passage-level retrieval, it is different with the datasets. We get the best performance when $k_{1}$ equals 100 in NQ, 500 in TriviaQA, 500 in WebQ, and 300 in the TREC dataset. 

\begin{table}[tp]
\centering
\resizebox{0.47\textwidth}{!}{
\begin{tabular}{lcccc}
\toprule
        & \textbf{NQ}  & \textbf{TriviaQA} & \textbf{WebQ} & \textbf{TREC} \\ \hline
DPR     &  75.5ms   &  86.5ms        &  78.5ms    &  91.5ms    \\
DHR-D   &  16.3ms   &   19.4ms      &  17.3ms    &  19.6ms    \\
DHR-P   &  2.5ms   &    9.9ms      &  7.2ms    &  4.5ms    \\ \hline
Speedup &  \textbf{4.02x}   &  \textbf{2.94x}         &  \textbf{3.20x}    &  \textbf{3.80x}   \\
\bottomrule
\end{tabular}
}
\caption{The comparison of retrieval time efficiency between DPR and the proposed DHR.}
\label{Table_run_time}
\end{table}

\subsection{Retrieval Time Efficiency}
During inference, DPR needs to search the gold passage from the 21-million passages.
In contrast, DHR only targets 5.38-million documents and the passages from retrieved top-$k_{1}$ documents. As discussed in the previous Section \ref{sen_anly}, $k_{1}$ is usually a small number like a few hundred. Therefore, the total amount of searching space decreases from 21-million to around 6-million. 

Since document embeddings and passage embeddings are encoded once after the model is trained, so we only discuss the index search time here. We run the best model of the first iteration on the test set twice and calculate the average index search time. We separately present the time cost on the document-level retrieval from the whole document corpus (shown in line DHR-D) and the time cost on the passage-level retrieval from the passage corpus of the retrieved documents (shown in line DHR-P) in Table \ref{Table_run_time}. The total time cost of our method is the addition of the DHR-D and DHR-P phrase time cost. And compared with the time cost in DPR, our proposed approach is nearly 3 to 4 times faster. This is a notable advantage of our method in practice.

\subsection{End-to-end QA System} \label{extractive_qa}
To test the end-to-end QA performance, we follow the DPR use extractive reader constructed by BERT. Given the top k retrieved passages (maximum 100 in our experiments), we combine the passage title, passage token with a special token \texttt{[SEP]} and send it to the reader. The reader assigns a passage selection score to each passage. In addition, it extracts an answer span from each passage by determining the start and end indexes and assigns a span score. The best span from the passage with the highest passage selection score is chosen as the final answer. And we declare the implementation detail in Appendix \ref{apx_endtoend}.  

Table \ref{Table_QA} shows our final end-to-end QA results compared with ORQA~\cite{lee2019latent} and DPR~\cite{karpukhin2020dense}, measured by exact match with the reference answer.
Overall, DHR leads to improvement of the QA scores on all four datasets.
For reference, we also experiment our retriever with a generative reader in Appendix \ref{sec_gen_reader}.

\begin{table}[tp]
\centering
\resizebox{0.49\textwidth}{!}{
\begin{tabular}{lcccc}
\toprule
\textbf{Model} & \textbf{NQ} & \textbf{TriviaQA} & \textbf{WebQ}  & \textbf{TREC} \\ \hline
BM25+BERT      &  26.5  &   47.1       &  17.7  & 21.3  \\
ORQA          &  33.3  &   45.0       &  36.4  & \textbf{30.1}  \\
DPR       &  41.5  &    56.8      &  34.6  & 25.9   \\
DPR*      &  42.4  &    56.9      & 35.5   & 26.0   \\\hline
DHR      &  \textbf{43.6}  &    \textbf{57.0}      & \textbf{36.6}   & 27.3  \\ 
\bottomrule
\end{tabular}
}
\caption{End-to-end QA (Exact Match) accuracy. The first block of results are copied from their cited papers.}
\label{Table_QA}
\end{table}
\section{Related Work}
\noindent{\textbf{Hierarchical Retrieval.}}
Hierarchical sparse retriever got attention in early 2000s.
\citet{levinson1992multilevel} proposed a multi-level hierarchical retrieval method in database search of conceptual graphs. In Web search, \citet{cui2003hierarchical} developed a structured document retriever which exploits both content and hierarchical structure of documents, and returns document elements with appropriate granularity.
\citet{bonab2019incorporating} incorporated hierarchical domain information into information retrieval models such that the domain specification resolves the ambiguity of questions.
Recently, \citet{nie2019revealing,asai2019learning} proposed a hierarchical retrieval approach with both paragraph and sentence level retrievers to extract supporting facts for the large-scale machine reading task. 

\noindent{\textbf{Dense Retrieval with Pre-trained Encoders.}}
With the strong embedding-based ability of the pre-trained model, \citet{lee2019latent,chang2020pre} showed the advantage of dual encoder framework with a set of pre-training tasks \citep{liu2020kg} can achieve strong baselines in the large-scale question-document retrieval task. DPR \citep{karpukhin2020dense} developed a better training scheme using contrastive learning and shows that without the pre-training task, just using a small number of training pairs can achieve state-of-the-art. DPR has been used as an important module in very recent works. \citet{xiong2020answering} extended the DPR to the multi-hop setting \citep{liu2020interpretable} and shows that DPR using passage text only to retrieve multi-hop passages can achieve good performance, without the help of the hyperlinks. 

Recent research explored various ways to construct better negative training instances for dense retrieval. ANCE \citep{xiong2020approximate} used the retrieval model trained in the previous iteration to discover new negatives and construct a different set of examples in each training iteration. \citet{lu2020neural} explored different types of negatives and uses them in both the pre-training and fine-tuning stages. 
The other direction of recent research works on improving the training strategy in dense retrieval. Rather than using the gold document as distant supervised training of retrieval, \citet{izacard2020distilling} leveraged attention score of a reader model to obtain synthetic labels for the retriever. And \citet{sachan2021end} presented the end-to-end supervised training of the reader and retriever. Furthermore, \citet{mao2020generation} generated various contexts of a question to enrich the semantics of the questions is beneficial to improve DPR retrieval accuracy. \citet{xiong2020progressively} used a pretrained sequence-to-sequence model to generate question-passage pairs for pretraining and proposed a simple progressive pretraining algorithm to ensure the effective negative samples in each batch. A pretrained sequence-to-sequence model is exploited to create question-passage pairs in the zero-shot setting \citep{ma2021zero}.

\section{Conclusion}
In this work, we propose Dense Hierarchical Retrieval (DHR) for open-domain QA and demonstrate that the hierarchical model provides evident benefits in terms of accuracy and efficiency. 
The hierarchical information is crucial to associate passages with documents such that the passage-level retriever tends to deviate from a misguided embedding function. 
Contrastive learning using proposed negatives further encourages a robust decision boundary between positives and hard negatives, leading to a meaningful fine-grained retriever. Extensive experiments and analysis on four Open-domain QA benchmarks demonstrate the effectiveness and efficiency of our proposed approach. 

\section*{Acknowledgements}
We would like to thank all the reviewers for their helpful comments. This work is supported by NSF under grants III-1763325, III-1909323,  III-2106758, and SaTC-1930941.

\clearpage

\begin{appendices}

\section{DHR Implementation Detail} \label{sec_retrieval_implementation}
Our document-level and passage-level retrievers use the base version of BERT as the pre-trained encoder. In the document-level retriever, we use the fixed token length 512 for document input and in the passage-level retriever, we use the fixed token length 280 for passage input. And the length of the question for both retrievers is 80. Our model is trained using the in-batch negative setting with a batch size of 128. We trained the document-level retriever and passage-level retriever for up to 40 epochs for large datasets (NQ, TriviaQA) and 100 epochs for small datasets (WebQ, TREC), with a learning rate of $10^{-5}$ using Adam, linear scheduling with warm-up and dropout rate 0.1. All the experiments are implemented on 8 A100 GPUs.

\section{Case study of DHR}
We compare the Top-1 retrieved passages and their corresponding documents from DHR and DPR. Table \ref{Table_case} shows an example that DHR retrieves the gold passages but DPR fails. The question asks for the person proposing the first DNA accurate model. The passage retrieved by DPR is under the section of History in the document DNA, which is relevant to the question but it doesn't contain the answer. The reason is mainly that the question asks for a DNA model rather than DNA. In contrast, the retrieved passage by DHR is under the topic of the DNA model and it contains the answer. It's hard for the dense retriever to retrieve the correct passage directly since the passage under DNA, history is so related to the question. But own to the help of the Document-level retrieval in our hierarchical retriever framework, it's easy to discover that the document DNA sequencing is much more related than the document DNA to the question.  

\begin{table*}[tp]
\centering
\resizebox{0.99\textwidth}{!}{
\begin{tabular}{l|l|l}
\textbf{Question} & \multicolumn{2}{l}{\textbf{Who proposed the first accurate model of DNA?} ~~~~~ \textbf{Answer: James Watson} }\\ \toprule
& \textbf{Retrieved Passage}& \textbf{Document of that Passage}\\ \hline
DPR  & \begin{tabular}[c]{@{}l@{}} \textbf{Title:} DNA\\ \textbf{DNA} was \textbf{first} isolated by the Swiss physician Friedrich Miesch-\\er who, in 1869, discovered a microscopic substance in the pus\\of discarded surgical bandages. As it resided in the nuclei of cells,\\he called it "nuclein". In 1878, Albrecht Kossel isolated the non-\\ protein component of "nuclein", nucleic acid, and later isolated\\its five primary nucleobases. In 1909, Phoebus Levene identified \\ the base, sugar, and phosphate nucleotide unit of the RNA (then \\ named "yeast nucleic acid"). In 1929, Levene identified deoxy-\\ ribose sugar in "thymus nucleic acid" (\textbf{DNA}).\\\textbf{Title list:} History\end{tabular} & \begin{tabular}[c]{@{}l@{}}\textbf{Title}: DNA \\ 
Deoxyribonucleic acid (; \textbf{DNA}) is a molecule composed of\\ 
two chains that coil around each other to form a double\\
helix carrying  the genetic instructions used in the growth, \\
development, ... The nitrogenous bases of the two of the \\ 
two separate polynucleotide strands are bound together,\\
according to base pairing rules, with hydrogen bonds to\\ 
make double-stranded \textbf{DNA}. \\
\textbf{Title list:} Properties, Nucleobase classification, Non-can-\\onical 
bases, Listing of non canonical bases found in DNA, \\ 
Base pairing, Sense and antisense, Supercoiling,...\end{tabular} \\ \hline
DHR  & \begin{tabular}[c]{@{}l@{}} \textbf{Title:} DNA sequencing\\ Deoxyribonucleic acid (\textbf{DNA}) was \textbf{first} discovered and isolated\\ by Friedrich Miescher in 1869, but it remained understudied for\\ many decades because proteins, rather than \textbf{DNA}, were thought \\ to hold the genetic blueprint to life .. This was the \textbf{first} time that\\ \textbf{DNA} was shown capable of transforming the properties of cells. \\ In 1953, \textcolor{red}{James Watson} and Francis Crick put forward their dou-\\ble-helix \textbf{model} of \textbf{DNA}, based on crystallized X-ray structures\\being studied by Rosalind Franklin 2013. \\ \textbf{Title list:} History, Discovery of DNA structure and function\end{tabular} & \begin{tabular}[c]{@{}l@{}} \textbf{Title:} DNA sequencing\\ \textbf{DNA} sequencing is the process of determining the order of\\ nucleotides in \textbf{DNA}. ...The advent of rapid \textbf{DNA} sequenc-\\ing methods has greatly accelerated biological and medical\\research and discovery. ... The \textbf{first} \textbf{DNA} sequences were ob-\\tained in the early 1970s by academic researchers using labor-\\ious methods based on two-dimensional chromatography.
  \\ \textbf{Title list:} Applications, ... History, Discovery of DNA struc- \\
ture and function, RNA sequencing, Early DNA sequencing\\
 methods, Sequencing of full genomes,..., Basic methods …
\end{tabular}                  
\end{tabular}
}
\caption{An example of passages returned by DPR and DHR and their corresponding document abstract. The words in bold means it appears in the question and the correct answers are written in \textcolor{red}{red}.}
\label{Table_case}
\end{table*}

\begin{table}[tp]
\centering
\resizebox{0.28\textwidth}{!}{
\begin{tabular}{l|cc}
\toprule
              & EM    & F1    \\ \hline
\textit{1-iter}        & & \\
DPR            & 48.20 & \\ 
DPR*           & 48.72 & 56.64 \\
DHR w/o Title & \textbf{50.63} & \textbf{58.74} \\
DHR w Title   & 49.86 & 57.97 \\ \hline
\textit{2-iter}        & & \\
DPR*           & 48.55 & 56.34 \\
DHR w/o Title & \textbf{50.33} & \textbf{58.28} \\
DHR w Title   & 50.27 & 58.20 \\
\bottomrule
\end{tabular}
}
\caption{End-to-end QA evaluation results on NQ test set using Fusion-in-Decoder model \cite{izacard2020leveraging}. * represents reproducing results on our processed Wikipedia data.}
\label{Table_FID}
\end{table}
\section{End-to-End QA} 
\subsection{Extractive Reader Implementation} \label{apx_endtoend}
For the implementation of the extractive reader, we sample 1 positive and 24 negative passages from the top 100 retrieved passages for each question. The training objective is to maximize the marginal log-likelihood of all the correct answer spans in the positive passage, combined with the log-likelihood of the positive passage being selected. We use the batch size of 16 for large datasets (NQ and TriviaQA) with a maximum of 40 epochs for large and 4 for small (WebQ and TREC) datasets with a maximum of 100 epochs. And we evaluate the development set at every 1000 steps. 
\subsection{QA results with the Generative Reader} \label{sec_gen_reader}
We implement our retrieval results on NQ test set with the Fusion-in-Decoder model (FiD) \citep{izacard2020leveraging}, a generative reader using pre-trained sequence-to-sequence model T5 \citep{raffel2019exploring}. The model takes the question, retrieved passages as input, and generates the answer. More precisely, each retrieved passage and its passage title are concatenated with the question and processed independently from other passages by the encoder. And the decoder calculates the attention over the concatenation of the joint representations of all the retrieved passages. 

We use top-50 retrieved passages for both training and inference, while T5-base is used as the underlying architecture. We train the model for 10 epochs with a batch size of 64 and a learning rate of $1e-4$. We evaluate the model on the development set at every 500 steps, and select the checkpoint obtaining the highest EM score as the final model, and report its results on the NQ test.

From the Table \ref{Table_FID}, we can see that our proposed model DHR outperforms the DPR results in both first and second iteration, even with the less retrieved passages (FiD implementation uses top-100 retrieved passages), which shows the better retrieval results lead to the better generative answering results. And the generative greatly outperforms the extractive approach in Section \ref{extractive_qa}. 

\end{appendices}

\end{document}